# Disentangling the components of a multiconfigurational excited state in isolated chromophore

Rodrigo Cezar de Campos Ferreira[1,3], Amandeep Sagwal[1,2], Jiří Doležal[1,3], Tomáš Neuman[1]*, and Martin Švec[1,3]*

[1] Institute of Physics, Czech Academy of Sciences; Cukrovarnická 10/112, CZ16200 Praha 6, Czech Republic

[2] Faculty of Mathematics and Physics, Charles University; Ke Karlovu 3, CZ12116 Praha 2, Czech Republic

[3] Institute of Organic Chemistry and Biochemistry, Czech Academy of Sciences; Flemingovo náměstí 542/2. CZ16000 Praha 6, Czech Republic

Abstract:

Studying the excited states of doublets is challenging for their typically multiconfigurational character. We employ light-scanning-tunneling microscopy (light-STM) to investigate photon-induced currents on a single open-shell PTCDA anion molecule placed into a plasmonic nanocavity between a tip and a substrate, irradiated by laser. Submolecular mapping reveals a zero-bias bidirectional photocurrent strongly varying with the lateral position of the tip apex above the molecule. We elucidate the mechanism in terms of a theoretical model in which a multiconfigurational doublet state is excited and decays back to the anion ground state through sequential electron transfers with the tip and the substrate. The correspondence of the experimental and theoretical contrast proves the correlated character of the excited state which can be described as a superposition of two dominating electronic configurations. By applying bipolar voltage on the junction with the molecule, we switch the dominant recombination pathway from one of the configurations to the other, effectively disentangling the multiconfigurational state individual components through visualization of their Dyson orbitals, as corroborated by theoretical modelling.

Doublet radical emitters are potential candidates for highly efficient optoelectronics as they eliminate inefficiencies connected to spin statistics of conventional closed-shell systems.[1,2] Their excited state can often be formed by a superposition of a few configurations involving the occupations of *different* molecular space orbitals (MOs), reflecting the simultaneous excitation of *different* electron-hole pairs.[3-11] The interference of these configurations can then result in a strong reduction or enhancement of optical transition dipoles and thus absorption and emission intensities related to the molecular electronic transition and is expected to impact the molecular optoelectronic properties. In particular, the electrons and holes emerging in the process of electron-hole pair splitting can originate from different electronic configurations, and hence from different pairs of occupied and unoccupied MOs, both belonging to the same excited state. This may lead to unintuitive dynamics of exciton splitting, governing e.g. the charge transfer in donor-acceptor interfaces in organic solar cells,[12] whose understanding relies on theoretical modelling,[13-16] but is hardly accessible by experimental methods. The widely employed angle-resolved photoelectron spectroscopy,[17] working on large molecular ensembles can solely provide averaged electron emission spectra that reflect the multiconfigurational character of excited states indirectly. Instead, a direct real-space observation of the orbitals involved in charge transfer upon photoexcitation of a single open-shell emitter made with submolecular resolution would be required.

To reach this goal, we employ the scanning tunneling microscopy (STM) with light to induce and detect photocurrent with atomic-scale precision.[18-20] This technique was applied to visualize photon-induced in-gap electron transport in a single closed-shell charge-neutral phthalocyanine [18] and pentacene [19] to elucidate how the frontier orbitals are involved in the transport-mediated decay of the optically excited states. The experiments confirmed the participation of the highest occupied MO (HOMO) and lowest unoccupied MO (LUMO) in the photocurrent channels as the optically excited states in the studied molecules have the HOMO-LUMO character.

Using this technique and complementing it with tip-enhanced photoluminescence (TEPL) mapping we unravel the multiconfigurational character of a doublet excited state of a singly-negatively-charged perylenetetracarboxylic dianhydride (PTCDA). Based on the acquired TEPL data, analysis of photocurrent maps, scanning-tunneling spectra and detailed theoretical modelling of the photon-driven process, we establish the essential electronic dynamics and composition of the excited state. We explore the sensitivity of the technique to the components constituting the superposition by controlling recombination channels, employing electrical field and tip position.

The concept of the simultaneous photocurrent and TEPL measurement with submolecular resolution is shown in Fig.1a. A STM Ag-tip is scanned over PTCDA molecules deposited on a decoupling layer of 3ML-NaCl on a crystalline Ag(111) surface, and the tunneling current and photon emission rate are recorded, using a standard high-gain preamplifier and a single-photon detector, respectively. The tip apex is illuminated by a laser spot of approximately 10 micrometer radius. Using the resonant antenna effect of the tip, the laser electric field is directed into the tip-surface gap plasmonic nanocavity and concentrated around the atomic-scale tip apex, where it interacts with the molecular excitations.[21-23] The excitation laser wavelength is above the energy needed to reach the first excited state of the PTCDA anion, the known equilibrium charge state in this scenario.[4,24,25] Moving the tip along the surface varies the coupling strength of the nanocavity with the molecules as well as the tunneling probability, and leads to tip-position dependent absorption, emission and charge transport.

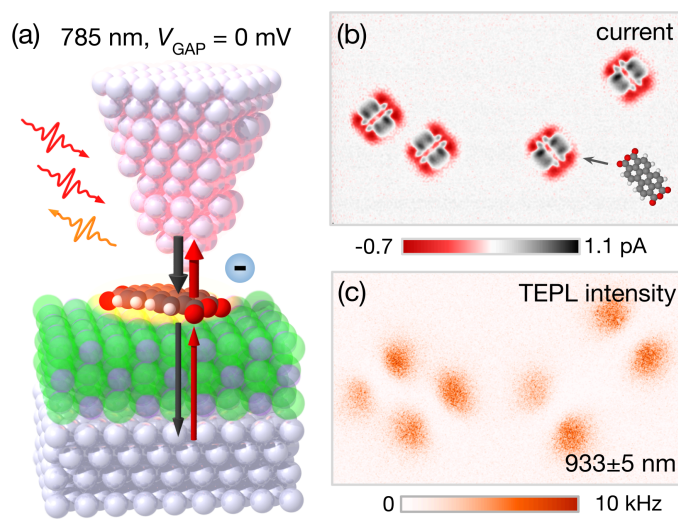

**Fig.1: Photon-induced current and luminescence of a single open-shell molecule in a STM-nanocavity.** (a) Scheme of the experiment: plasmonic tip apex is illuminated by focused laser light and scanned parallel to the surface plane of 3ML-NaCl/Ag(111) above an organic radical dye, PTCDA anion. The laser light is focused by the plasmonic cavity and excites the molecule, inducing photoluminescence and electron tunneling between the molecule and the tip/sample. (b) The recorded net tunneling current map above a group of four identical molecules. A model of PTCDA is shown for orientation (not to scale). (c) A map of the photon rate at the photodetector with a 920-945 nm filter. The measurements were taken with 785 nm excitation wavelength and at zero bias voltage. The map sizes are 14 x 8.5 nm$^2$.

Figs.1b,c show the recorded spatially-resolved zero-bias tunneling current and the photon rate at the energy of the fundamental PTCDA anion emission above an area with four identical molecules in two orientations with respect to the lattice of the NaCl. The magnitude and direction of the net current varies strongly with the precise location of the tip above their backbones, changing from negative (red) to positive (grey) at distances comparable to bond lengths and reaching pA values. Since the applied bias voltage is set to zero, the current is solely driven by the absorption of the incident light. Without the laser illumination, no current is observed at zero bias (see Fig.S6). The complex character of the contrast is strikingly different from the one observed on the same system without irradiation at positive-ion-resonance (PIR) and negative-ion-resonance (NIR) onsets where it is dominated by the singly occupied and singly unoccupied MOs (SOMO and SUMO).[4,26,25] This hints at the involvement of other frontier orbitals in the recombination channels at play.[27] The TEPL maps for the same area taken at the energy of luminescence of the PTCDA anion first excited state show two-lobe patterns above each molecule, oriented in the direction of the PTCDA longitudinal axes. It should be noted that the detected TEPL emission patterns in fact reflect the combined probabilities of absorption and emission, both dependent on the nanocavity position.[22,28,29]

An interpretation of the observed characteristic contrast in the photon-induced current maps requires identifying the states of the molecule and relevant charge transport pathways that may significantly contribute to the net current. To that end we develop a model considering the most relevant molecular states and transitions among them, capturing the essential mechanisms involved in the formation of the photocurrents and TEPL under the explored experimental conditions (schematically depicted in Fig.2a, details in SI). The $D_0^- \leftrightarrow D_1^-$ and $D_0^- \leftrightarrow D_2^-$ optical gaps are apparent from the TEPL spectra (Fig.2b,c), assuming negligible Stokes shifts between the absorption and emission in our scenario; the neutral $S_0$ and doubly negative $S_0^{2-}$ levels and the effect of substrate reorganization (represented by the gray shading above the ground-state levels of the three charge states) are deduced from matching the electron transport resonances measured in d$I$/d$V$ spectroscopy without illumination (dark current in Fig.2d).[4,25] These levels are consistent with recent single-electron charging experiments (see Table S1).[24,30,31] Time-dependent density-functional theory (TD-DFT) simulations provide the respective electronic configurations for each of the levels that can be represented by the frontier orbital occupations. In this model, the driving mechanism is the absorption of a photon, mediated by the plasmonic cavity, converting the $D_0^-$ state into the $D_1^-$ or $D_2^-$ excited states. The experimental evidence based on the analysis of TEPL patterns discussed further below points toward a mechanism where either a vibronic Franck-Condon $D_0^- \to D_1^-(\nu_i)$ or a Herzberg-Teller $D_0^- \to D_2^-(\nu_j)$ transition is induced and the excitation is subsequently distributed between the $D_1^-$ and $D_2^-$ states by the vibrationally driven internal conversion. In both cases, the spatial distribution of the absorption is driven by the $D_0^- \to D_1^-$ zero-phonon transition density.

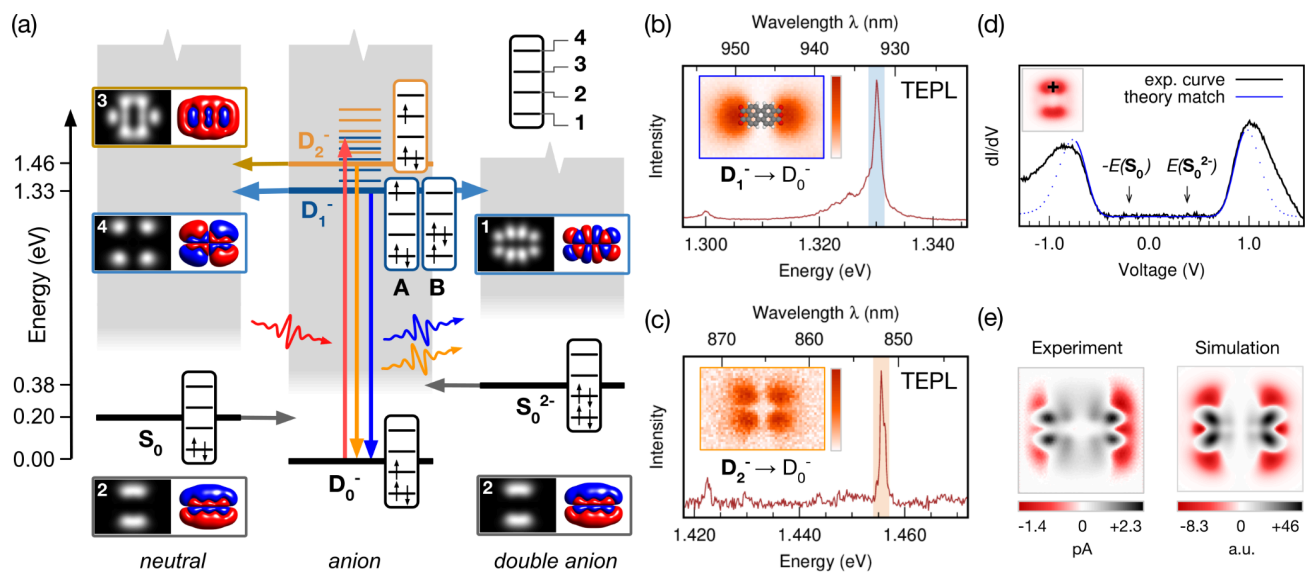

**Fig.2: The recombination pathways of the PTCDA anion under irradiation.** (a) Scheme of the excitations and transitions (denoted by vertical and horizontal arrows, respectively) between the five considered energy states of the molecule with various charges and electronic configurations, at zero bias. $S_0$ is the neutral ground state, $D_0^-/D_1^-/D_2^-$ are the singly-negative charged ground and first two excited states, respectively. The $S_0^{2-}$ is the doubly-negative charged ground state. The state $D_1^-$ is composed of two dominating configurations denoted as A and B, all the other states are well-described as single-configurational states. The gray band above each state represents the effect of a threshold function reflecting the substrate reorganization energy. The legend identifies the orbitals involved in the transitions, which are shown in the insets with the corresponding simulated spatially-dependent probability of tunneling. (b,c) TEPL spectra and the corresponding maps of the $D_1^-$ and $D_2^-$ radiative decays derived from TEPL mapping measurements. Dimensions of both maps are 5 x 3.5 nm². (d) Matching of the resonant onsets in the d$I$/d$V$ measured on a PTCDA (location in the inset) without illumination to estimate the $S_0$, $S_0^{2-}$ and substrate reorganization energies, using the calculated threshold function. The solid line denotes the matched part of the theoretical curve. (e) The comparison of the experimental and simulated photon-induced tunneling current.

Theoretical calculations show the multiconfigurational character of the doublet $D_1^-$ state which is mainly composed of two electronic configurations, denoted A and B in the scheme of Fig.2a. Consequently, *two* major indirect decay paths are allowed from the $D_1^-$ to the $D_0^-$ ground state, each involving two single-electron resonant-charge-transfer steps (horizontal arrows) *via* one of the singlet $S_0$ or $S_0^{2-}$ states. The steps are realized by sequentially giving up an excited electron from the molecule and filling the hole by the resonant charge transfer from the substrate or the tip (the $D_1^- \rightarrow S_0 \rightarrow D_0^-$ path), or *vice versa* - filling the hole first and then giving up the excess electron (the $D_1^- \rightarrow S_0^{2-} \rightarrow D_0^-$ path). In contrast, the $D_2^-$ state has a single-configurational character and, in the first step, is only allowed to decay by a resonant electron transfer into the $S_0$ state. The path through $S_0^{2-}$ involving an electron capture is unlikely due to the mismatch of the electron configurations of the $D_2^-$ and $S_0^{2-}$ states as the configuration of $S_0^{2-}$ cannot be obtained by adding a single electron into the configuration of $D_2^-$. Within the framework of this simplified picture, in which the neutral or doubly negative triplet states, photon-assisted tunneling, and role of hot electrons are omitted,[32] we can identify and computationally evaluate two simultaneously acting mechanisms that spatially modulate the observed photon-induced current depending on the tip-position: (i) tunneling probabilities and (ii) plasmon-exciton coupling.

The first one arises from the spatial modulation of the tip-molecule tunneling probability. The photocurrent is nonzero only if the tunneling process results in a net transfer of charge from the tip to the substrate or vice versa, *i.e.*, the tunneling between the tip and the molecule is followed or preceded by a transfer from/to the substrate. The presence of nonzero photocurrent therefore always involves the tip-molecule tunneling step, which is tip-position-dependent and scales with the density of states (DOS) derived from the molecular Dyson orbital related to the respective tip-mediated tunneling event.[33-35] This DOS can be approximated as the square modulus of the MOs, which are either being populated or depopulated, evaluated at a constant height in a close distance to the molecule (the grayscale insets 1-4 of the Fig.2a, details of the calculations are provided in the SI).

The second mechanism modulates the contrast as a function of the tip position on a few-angstrom scale characteristic for the nanocavity confinement.[36] The modulation originates from the tip-position-dependent absorption that proportionally determines the excitation rate of $D_1^-$, as inferred in the recent work of Zhu *et al.*,[19] but also from the effect of tip-position-dependent emission probability that competes with the non-radiative decay paths and co-determines the $D_1^-$ equilibrium population. The respective spatial dependencies can be deduced from the patterns of the emission from the $D_1^-$ and $D_2^-$ in TEPL maps in Figs.2b,c. For the $D_1^-$ (emission energy 1.330 eV), it is the characteristic two-lobe pattern oriented along the longitudinal axis of the molecule, which can be attributed to the sequence of absorption, thermalization and emission, *i.e.* $D_0^- \rightarrow D_1^-(v_i)/D_2^-(v_j) \rightarrow D_1^- \rightarrow D_0^-$, where $D_1^-(v_i)$ and $D_2^-(v_j)$ are the vibronically excited states. On the other hand, the four-lobe intensity map of the $D_2^-$ emission (at 1.456 eV) is strikingly different from the expected two-lobe pattern oriented along the short axis of the molecule.[4] Our interpretation is that the $D_2^-$ is reached indirectly, through an internal conversion from the vibronically excited state of $D_1^-$, by undergoing the $D_0^- \rightarrow D_1^-(v_i)/D_2^-(v_j) \rightarrow D_2^- \rightarrow D_0^-$ process, in which the spatially-dependent probabilities of $D_0^- \rightarrow D_1^-$ and $D_2^- \rightarrow D_0^-$ transitions multiply and generate the four-lobed pattern (a simulation is shown in Fig.S7). Consequently, both absorption probabilities as well as the emission follow the tip-position dependent nanocavity coupling with the $D_0^- \leftrightarrow D_1^-$ transition, which we simulate using the transition charge density from TD-DFT convolved with the function representing the plasmonic quasi-electrostatic potential of the nanocavity.[4,37]

To simulate the current and TEPL maps, we calculate the rates associated with the position-dependent plasmon-exciton interaction and all considered charge-exchanges between the PTCDA, the tip and the substrate and use them to create a set of rate equations (see the SI). The solution yields the spatially-dependent net current flowing between the tip and the molecule that reproduces the observed contrast, as shown by the comparison with the experiment in Figs.2e. Because the tunneling probabilities are evaluated in a plane close to the molecule, the orbital-like character of the maps has a high spatial resolution.[38] The current maps calculated using the zero-bias model indicate that the three transitions from the $D_1^-$ and $D_2^-$ to $S_0$ and $S_0^{2-}$ that involve orbitals 1, 3 and 4 play a similarly important role in the experimentally observed contrast. On the other hand, the SOMO- and SUMO-like character associated with the transitions of $S_0$ and $S_0^{2-}$ to $D_0^-$, reflecting the shape of the orbital 2, is less apparent. The decay channels involving $D_1^-, D_2^- \rightarrow S_0$ or $D_1^- \rightarrow S_0^{2-}$ contribute with a characteristic pattern that is a fingerprint of the respective electron configuration forming the excited states. The positive current (mapped to gray color scale in Fig.2e) is dominantly generated by the injection of an electron into the unoccupied orbital 1 of the configuration *B* of $D_1^-$. The negative portion of the current (red color scale) is mostly generated by two contributions. One involves the extraction of an electron from orbital 3 that is occupied in the configuration of $D_2^-$ and the second one is associated with the extraction of an electron from the occupied orbital 4 in configuration *A* of the $D_1^-$ state (see the orbital occupations in Fig.2a for reference). The signatures of both orbital 1 (in the positive current) and orbital 4 (in the negative current) in the photon-induced current map therefore prove the presence of both configurations A and B in the state $D_1^-$.

We can test the possibility of disentangling different components of the multiconfigurational excited state by selectively promoting the decay path involving *either the A- or the B-configuration* of $D_1^-$. To that end, we tune the bias voltage and anticipate one of the two patterns that identifies the respective pathway in the current maps. To separate the photon-induced component from the total tunneling current (*I*) we modulate the laser intensity and detect the response in the current - obtaining the d*I*/d*P* signal, which is proportional to the photocurrent in our scenario (see Fig.S5 for details of the setup).

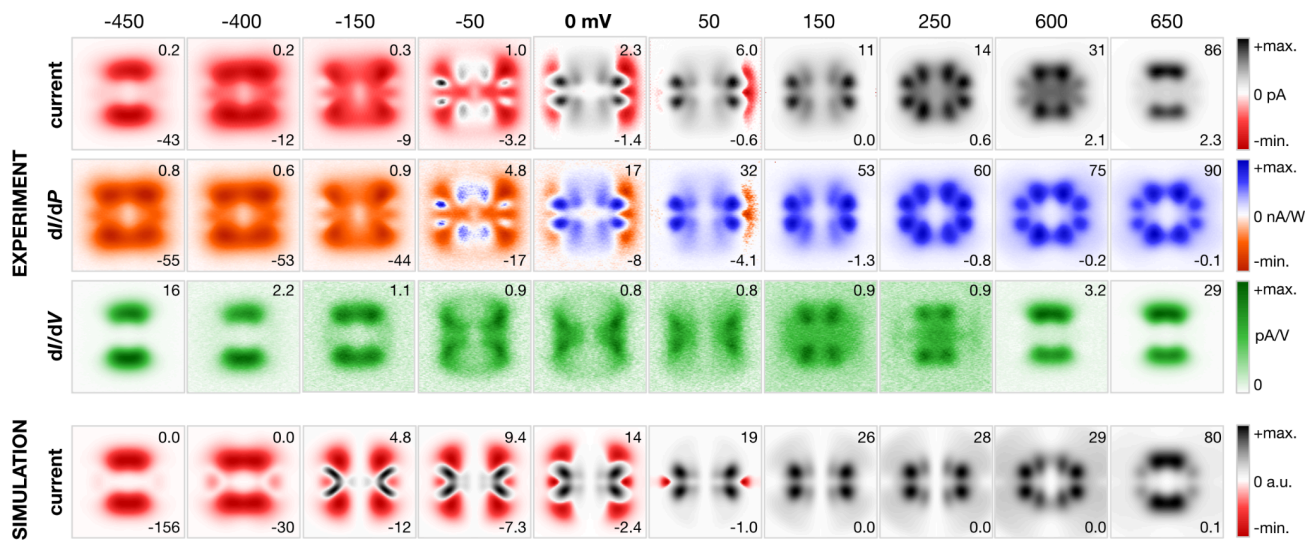

**Fig.3: Comparison of bias-dependent maps of current, photocurrent, d*I*/d*V* and simulations.** The current, the photon-induced component of the current proportional to photocurrent (d*I*/d*P*) and conductance (d*I*/d*V*) maps are taken simultaneously over a PTCDA molecule in a constant-height mode as a function of the applied bias voltage in the range from -450 to 650 mV. Theoretical simulations are provided for comparison. The area of the images is 2 x 2 nm$^2$. The numbers at the top and bottom right corner of each image represent the minimum and maximum data value, respectively. The "+max" and "-min" in the color scale bar refer to the maximum positive and minimum negative intensity of the signal.

The detailed maps of the bias-dependent current, d*I*/d*P* and the differential conductance d*I*/d*V* under irradiation are shown in Fig.3. The bidirectionality of the net electron transport over the PTCDA molecule is gradually lost as the bias is increased to higher positive or negative values. At positive bias above 150 mV the current map strongly resembles the tunneling probability distribution generated for orbital 1, corresponding to the $D_1^- \rightarrow S_0^{2-}$ transition, suggesting that at this bias the contrast is dominated by the process where an electron from the tip is injected into this orbital. Similarly, at the negative polarity below -150 mV, the current map is comparable to the tunneling from orbitals 3 and 4 ($D_2^-, D_1^- \rightarrow S_0$). At both onsets of the resonant tunneling, -450 and 600 mV, the current is apparently dominated by the extraction from SOMO and injection to SUMO, respectively, which is detected also in the d*I*/d*V* maps. While the d*I*/d*P* map essentially corresponds to the current map recorded within the gap region (approx. -350 to 250 mV) it retains its character beyond the onsets of resonant tunneling, where it reaches a saturation. The same trend can be reproduced using the theoretical model that considers the bias, including a voltage drop between the molecule and the substrate. The voltage-induced changes of the photocurrent patterns can be understood as a result of the variation of the transition rates promoting either the $D_1^- \rightarrow S_0 \rightarrow D_0^-$ or the $D_1^-, D_2^- \rightarrow S_0^{2-} \rightarrow D_0^-$ path as a function of the bias voltage, which occurs due to the smooth onset of the thresholds of the tunneling steps $S_0 \rightarrow D_0^-$ and $S_0^{2-} \rightarrow D_0^-$ (see also Fig.S4 for more details).

The set of experimental d*I*/d*V* maps in Fig.3 measured within the transport gap shows an overall weak magnitude compared to the resonant onsets. Yet, the contrast is significantly more intense than the almost negligible signal measured without illumination (Fig.S6) and shows very specific patterns. This indicates that the conduction channels are modified in the transport gap as a result of the photon absorption and charge transfer decay of the excited states. The d*I*/d*V* curves measured at two positions over the molecule with the tip close to the molecule in Fig.S8 reveal relatively broad peaks near to zero bias depending on the measurement location. We observe a single peak in the location corresponding to the outer lobes of the orbital 1 and a doublet of peaks above the inner lobes. The d*I*/d*V* maps and curves extracted from the results of the model qualitatively reproduce this behavior, confirming its validity.

In conclusion, by imaging photocurrents generated in a single open-shell PTCDA anion molecule using the light-STM, we have identified signatures of the two dominating configurations composing its multiconfigurational excited state $D_1^-$. We found that these two configurations are fingerprinted in the tip-position-dependent photocurrent maps as a contrast associated with positive and negative photocurrents, respectively. This finding is supported by a theoretical model combining ab-initio calculations of molecular properties with rate-equations describing the charge-transfer, plasmon-enhanced photon absorption and emission. Furthermore, we have experimentally demonstrated that it is possible to achieve selective mapping of the photocurrent paths associated with the positive and negative current contributions. Such tuning is enabled by exploiting the smooth variation of the voltage-dependent tunneling rates between the molecule and the tip/substrate even at small bias voltages. Such smooth variation arises due to the substrate reorganization upon changing the molecular charge, combined with the relatively small energy of tunneling onsets associated with the alignment of the many-body states of PTCDA. Our findings prove that an optically-driven variety of the well-established STS mapping can be instrumental in deciphering the electronic character of hard-to-access multi-configurational excited states of molecules by directly imaging the constituent electronic configurations.

**Acknowledgements:**

We are grateful to Sofia Canola for insightful suggestions regarding the multiconfigurational states and Anna Roslawska for the support and remarks on the manuscript. R.C.C.F., M.Š., A.S. acknowledge the funding from the Czech Science Foundation grant no. 22-18718S and the support from the CzechNanoLab Research Infrastructure supported by MEYS CR (LM2023051) J.D. is grateful to the IOCB Postdoctoral fellowships programme. T. N. acknowledges the Lumina Quaeruntur fellowship of the Czech Academy of Sciences. Computational resources were supplied by the project "e-Infrastruktura CZ" (e-INFRA CZ LM2018140) supported by the Ministry of Education, Youth and Sports of the Czech Republic.

**Experimental methods:**

A fiber coupled excitation source (Thorlabs SHG stabilized laser diode) at 785 nm was used with power of 400 μW, modulated by 200 μW at 743 Hz. The bias modulation for the d$I$/d$V$ measurements was 10 mV at 923 Hz. Plasmonic tips with a spectrum matching the excitation and emission wavelengths were prepared by a series of controlled indentations into a clean Ag(111) surface. All presented experiments were done in the constant-height mode with a Createc SPM at a temperature of 5.6 K. The confocal optical setup for coupling the external laser was based on the layout described in previous works.[5,28] For the photon rate and TEPL spectra acquisition we used the PerkinElmer SPCM-AQR-15 single-photon avalanche detector and the Andor Kymera 328i with Newton 920 CCD (BEX2-DD sensor), respectively.

**References**

[1] X. Ai, E. W. Evans, S. Dong, A. J. Gillett, H. Guo, Y. Chen, T. J. H. Hele, R. H. Friend, and F. Li, Efficient radical-based light-emitting diodes with doublet emission, Nature 563, 536 (2018).

[2] C. Tonnelé and D. Casanova, Rationalization and tuning of doublet emission in organic radicals, J. Mater. Chem. C 10, 13826 (2022).

[3] A. Heil and C. M. Marian, DFT/MRCI-R2018 study of the photophysics of the zinc(ii) tripyrrindione radical: non-Kasha emission?, Phys. Chem. Chem. Phys. 21, 19857 (2019).

[4] J. Doležal, S. Canola, P. Hapala, R. C. de Campos Ferreira, P. Merino, and M. Švec, Real Space Visualization of Entangled Excitonic States in Charged Molecular Assemblies, ACS Nano 16, 1082 (2021).

[5] R. C. de Campos Ferreira, A. Sagwal, J. Doležal, S. Canola, P. Merino, T. Neuman, and M. Švec, Resonant Tip-Enhanced Raman Spectroscopy of a Single-Molecule Kondo System, ACS Nano 18, 13164 (2024).

[6] M. Gouterman, Spectra of porphyrins, Journal of Molecular Spectroscopy 6, 138 (1961).

[7] A. B. J. PARUSEL and S. GRIMME, DFT/MRCI calculations on the excited states of porphyrin, hydroporphyrins, tetrazaporphyrins and metalloporphyrins, J. Porphyrins Phthalocyanines 05, 225 (2001).

[8] K. Schulten, I. Ohmine, and M. Karplus, Correlation effects in the spectra of polyenes, The Journal of Chemical Physics 64, 4422 (1976).

[9] H. F. Bettinger, C. Tönshoff, M. Doerr, and E. Sanchez-Garcia, Electronically Excited States of Higher Acenes up to Nonacene: A Density Functional Theory/Multireference Configuration Interaction Study, J. Chem. Theory Comput. 12, 305 (2015).

[10] S. Jiang, T. Neuman, A. Boeglin, F. Scheurer, and G. Schull, Topologically localized excitons in single graphene nanoribbons, Science 379, 1049 (2023).

[11] A. Abdurahman, T. J. H. Hele, Q. Gu, J. Zhang, Q. Peng, M. Zhang, R. H. Friend, F. Li, and E. W. Evans, Understanding the luminescent nature of organic radicals for efficient doublet emitters and pure-red light-emitting diodes, Nat. Mater. 19, 1224 (2020).

[12] Z. Xu, Y. Zhou, C. Y. Yam, L. Groß, A. De Sio, T. Frauenheim, C. Lienau, and G. Chen, Revealing generation, migration, and dissociation of electron-hole pairs and current emergence in an organic photovoltaic cell, Sci. Adv. 7, (2021).

[13] M. K. Kristiansson et al., Experimental and theoretical studies of excited states in Ir$^-$, Phys. Rev. A 103, (2021).

[14] K. Regeta, C. Bannwarth, S. Grimme, and M. Allan, Free electrons and ionic liquids: study of excited states by means of electron-energy loss spectroscopy and the density functional theory multireference configuration interaction method, Phys. Chem. Chem. Phys. 17, 15771 (2015).

[15] B. Helmich-Paris, CASSCF linear response calculations for large open-shell molecules, The Journal of Chemical Physics 150, (2019).

[16] H. Lischka, D. Nachtigallová, A. J. A. Aquino, P. G. Szalay, F. Plasser, F. B. C. Machado, and M. Barbatti, Multireference Approaches for Excited States of Molecules, Chem. Rev. 118, 7293 (2018).

[17] D. Lüftner, T. Ules, E. M. Reinisch, G. Koller, S. Soubatch, F. S. Tautz, M. G. Ramsey, and P. Puschnig, Imaging the wave functions of adsorbed molecules, Proc. Natl. Acad. Sci. U.S.A. 111, 605 (2013).

[18] M. Imai-Imada et al., Orbital-Resolved Visualization of Single-Molecule Photocurrent Channels, Nature 603, 829 (2022).

[19] R. Zhu, X.-R. Dong, B. Yang, R.-L. Han, W.-J. Mao, G. Chen, Y. Zhang, Y. Zhang, and Z.-C. Dong, Revealing Single-Molecule Photocurrent Generation Mechanisms under On- and Off-Resonance Excitation, Nano Lett. (2024).

[20] M. Garg, A. Martin-Jimenez, Y. Luo, and K. Kern, Ultrafast Photon-Induced Tunneling Microscopy, ACS Nano 15, 18071 (2021).

[21] C. Chen, P. Chu, C. A. Bobisch, D. L. Mills, and W. Ho, Viewing the Interior of a Single Molecule: Vibronically Resolved Photon Imaging at Submolecular Resolution, Physical Review Letters 105, (2010).

[22] B. Yang et al., Sub-Nanometre Resolution in Single-Molecule Photoluminescence Imaging, Nature Photonics 14, 693 (2020).

[23] R. B. Jaculbia, H. Imada, K. Miwa, T. Iwasa, M. Takenaka, B. Yang, E. Kazuma, N. Hayazawa, T. Taketsugu, and Y. Kim, Single-Molecule Resonance Raman Effect in a Plasmonic Nanocavity, Nature Nanotechnology 15, 105 (2020).

[24] L. Sellies, J. Eckrich, L. Gross, A. Donarini, and J. Repp, Controlled Single-Electron Transfer Enables Time-Resolved Excited-State Spectroscopy of Individual Molecules, Nature Nanotechnology (2024).

[25] K. A. Cochrane, A. Schiffrin, T. S. Roussy, M. Capsoni, and S. A. Burke, Pronounced Polarization-Induced Energy Level Shifts at Boundaries of Organic Semiconductor Nanostructures, Nature Communications 6, (2015).

[26] K. Kimura, K. Miwa, H. Imada, M. Imai-Imada, S. Kawahara, J. Takeya, M. Kawai, M. Galperin, and Y. Kim, Selective Triplet Exciton Formation in a Single Molecule, Nature 570, 210 (2019).

[27] P. Scheuerer, L. L. Patera, and J. Repp, Manipulating and Probing the Distribution of Excess Electrons in an Electrically Isolated Self-Assembled Molecular Structure, Nano Letters 20, 1839 (2020).

[28] J. Doležal, A. Sagwal, R. C. de Campos Ferreira, and M. Švec, Single-Molecule Time-Resolved Spectroscopy in a Tunable STM Nanocavity, Nano Letters 24, 1629 (2024).

[29] A. Rosławska, K. Kaiser, M. Romeo, E. Devaux, F. Scheurer, S. Berciaud, T. Neuman, and G. Schull, Submolecular-scale control of phototautomerization, Nat. Nanotechnol. 19, 738 (2024).

[30] N. Friedrich et al., Fluorescence from a single-molecule probe directly attached to a plasmonic STM tip, Nat Commun 15, (2024).

[31] D. Hernangómez-Pérez, J. Schlör, D. A. Egger, L. L. Patera, J. Repp, and F. Evers, Reorganization energy and polaronic effects of pentacene on NaCl films, Phys. Rev. B 102, (2020).

[32] C. Lin, F. Krecinic, H. Yoshino, A. Hammud, A. Pan, M. Wolf, M. Müller, and T. Kumagai, Continuous-Wave Multiphoton-Induced Electron Transfer in Tunnel Junctions Driven by Intense Plasmonic Fields, ACS Photonics 10, 3637 (2023).

[33] J. V. Ortiz, Dyson-orbital concepts for description of electrons in molecules, The Journal of Chemical Physics 153, (2020).

[34] V. Pomogaev, S. Lee, S. Shaik, M. Filatov, and C. H. Choi, Exploring Dyson's Orbitals and Their Electron Binding Energies for Conceptualizing Excited States from Response Methodology, J. Phys. Chem. Lett. 12, 9963 (2021).

[35] D. G. Truhlar, P. C. Hiberty, S. Shaik, M. S. Gordon, and D. Danovich, Orbitals and the Interpretation of Photoelectron Spectroscopy and (e,2e) Ionization Experiments, Angew Chem Int Ed 58, 12332 (2019).

[36] T. Neuman, R. Esteban, D. Casanova, F. J. García-Vidal, and J. Aizpurua, Coupling of Molecular Emitters and Plasmonic Cavities beyond the Point-Dipole Approximation, Nano Letters 18, 2358 (2018).

[37] A. Rosławska, T. Neuman, B. Doppagne, A. G. Borisov, M. Romeo, F. Scheurer, J. Aizpurua, and G. Schull, Mapping Lamb, Stark, and Purcell Effects at a Chromophore-Picocavity Junction with Hyper-Resolved Fluorescence Microscopy, Phys. Rev. X 12, (2022).

[38] K. Kaiser, S. Jiang, M. Romeo, F. Scheurer, G. Schull, and A. Rosławska, Gating Single-Molecule Fluorescence with Electrons, Physical Review Letters 133, (2024).

# Supplementary Information
# Disentangling the components of a multiconfigurational excited state in isolated chromophore

Rodrigo Cezar de Campos Ferreira[1,3], Amandeep Sagwal[1,2], Jiří Doležal[1,3], Tomáš Neuman[1]*, and Martin Švec[1,3]*

[1] Institute of Physics, Czech Academy of Sciences; Cukrovarnická 10/112, CZ16200 Praha 6, Czech Republic

[2] Faculty of Mathematics and Physics, Charles University; Ke Karlovu 3, CZ12116 Praha 2, Czech Republic

[3] Institute of Organic Chemistry and Biochemistry, Czech Academy of Sciences; Flemingovo náměstí 542/2. CZ16000 Praha 6, Czech Republic

**Theoretical methods**

*Rate equations.* The theoretical model of the photocurrent generation relies on solving a system of rate equations governing the transitions among the many-body states of the PTCDA molecule as sketched in Fig. 2a. Steady-state solutions of the rate equations are obtained for each lateral position of the tip to generate the photocurrent maps. More explicitly, for each many-body state population $N_i$, where $i$ runs over the many-body states $S_0$, $D_0^-$, $D_1^-$, $D_2^-$, $S_0^{2-}$, we have a system of rate equations in the steady-state:

$$0 = \frac{dN_i}{dt} = -\sum_j (\gamma_{ji}^T + \gamma_{ji}^S + \kappa_{ji})N_i + \sum_j (\gamma_{ij}^T + \gamma_{ij}^S + \kappa_{ij})N_j, \qquad (1)$$

where the rates $\gamma_{ji}^T$ and $\gamma_{ji}^S$ are the tip and substrate mediated tunneling rates connecting state $i$ to state $j$ of different charge and $\kappa_{ji}$ are the plasmon-mediated excitation and deexcitation rates connecting the states of the same charge. We do not consider spin degeneracy, for simplicity. The tunneling currents are obtained as

$$I \propto \sum_{ij} N_i \gamma_{ji}^T (Q_i - Q_j), \qquad (2)$$

with $Q_i$ being the total charge of the molecular state $i$.

The tip-mediated tunneling rates $\gamma_{ji}^T$ used in the model carry both the spatial dependence and the bias-voltage ($V$) dependence and have the form:

$$\gamma_{ji}^T = \gamma_0^T DOS_{ji}(x,y)\Theta(E_i^T - E_j^T),\qquad(3)$$

where

$$DOS_{ji}(x,y) = |\psi_{ij}(x,y,z=z_T)|^2/max(|\psi_{ij}(x,y,z=z_T)|^2),\qquad(4)$$

with $\psi_{ij}(x,y,z=z_T)$ being an approximation of the Dyson orbitals of the transition from $i$ to $j$, evaluated at a plane $z = z_T$, $\Theta(E_i^T - E_j^T)$ being a threshold function, and $\gamma_0^T$ being a scaling constant. We use $\gamma_0^T = 1$ μeV as it simulates the situation when the tip-molecule tunneling is dominating over the substrate-molecule tunneling,[1] as expected in the experiment due to the small tip-molecule distance required to detect photocurrents. We calculate $\psi_{ij}(x,y,z=z_T)$ using the density-functional theory from which we extract the orbitals that correspond to the dominating single-particle change of configuration between many-body states $i$ and $j$, as described below. The threshold function takes as inputs the energies $E_i^T$ that are related to the many-body state energies $E_i$ and are evaluated assuming that the electron is exchanged between the molecule and the tip, and the voltage $V$. $E_i^T$ is related to $E_i$ as follows:

$$E_i^T = E_i + \alpha Q_i V,\qquad(5)$$

where $\alpha = 0.8$ accounts for the fraction of the voltage drop across the molecule-tip gap. The value of $\alpha$ is justified by the small tip-molecule distance and a good match of the simulation with experimental data, particularly the d$I$/d$V$ curves. The many-body energies $E_i$ are summarized in Table 1 and are derived from the experimentally obtained data including the photon emission energies and d$I$/d$V$ spectra. The procedure of obtaining these energies is described below in the next section.

The substrate-mediated rates $\gamma_{ji}^S$ are calculated similarly, but are independent of the lateral position of the tip:

$$\gamma_{ji}^S = \gamma_0^S \Theta(E_i^S - E_j^S),\qquad(6)$$

with $\gamma_0^S = 0.1$ μeV. Here the superscript $S$ indicates that the respective parameters are considered for the substrate side, whose values will generally differ for the ones on the side of the tip as they are dependent on the voltage drop between the molecule and the underlying metal:

$$E_i^S = E_i - (1-\alpha)Q_i V.\qquad(7)$$

The charge-conserving plasmon-mediated decay $D_1^- \to D_0^-$ and $D_2^- \to D_0^-$ are modelled assuming that they are mediated by the interaction of the transition density $\rho_{ji}$ corresponding to the respective transition from state $i$ to $j$ with the tip plasmon, whose quasi-electrostatic potential $\phi_{pl}$ is localized around the tip apex. The plasmon-mediated decay rates are then calculated as

$$\kappa_{ji} = \kappa_0 + \kappa_{pl} |g_{ji}(x,y)|^2 / max(|g_{ji}(x,y)|^2), \qquad (8)$$

with

$$g_{ji}(x,y) = \int \rho_{ji}(X,Y,Z) \phi_{pl}(X-x, Y-y, Z) d^3R, \qquad (9)$$

where $(X, Y, Z)$ are cartesian coordinates and the highest point of the molecule lies in the plane $Z = 0$. The parameters $\hbar\kappa_0 = 0.1$ meV and $\hbar\kappa_{pl} = 0.5$ meV account for the substrate-mediated tip-position-independent "background" component of the exciton decay (e.g. due to the electron-hole pair excitations in the substrate) and the tip-position-dependent decay, respectively, and their values are selected and modelled consistently with previous studies.[2] To model the plasmonic potential $\phi_{pl}$ we used the simple point-charge model

$$\phi_{pl}(X, Y, Z) = 1/\sqrt{X^2 + Y^2 + (Z-z_1)^2} - 1/\sqrt{X^2 + Y^2 + (Z+z_2)^2}. \qquad (10)$$

Here $z_1 = 30$ a.u. and $z_2 = 40$ a.u. are suitable chosen parameters approximating the localization of the plasmonic field.

To implement the excitation process, we choose for definiteness that the energy is absorbed via a Herzberg-Teller transition into $D_2^-$ (with transition dipole borrowed from $D_0^- \to D_1^-$) and is then redistributed by internal conversion (IC) between $D_1^-$ and $D_2^-$. The plasmon-mediated excitation process $D_0^- \to D_2^-$ is implemented similarly to the decay rates as:

$$\kappa_{D_2^- D_0^-} = \eta_0 |g_{D_1^- D_0^-}(x,y)|^2 / max(|g_{D_1^- D_0^-}(x,y)|^2), \qquad (11)$$

where $\hbar\eta_0 = 3\times 10^{-7}$ represents a weak pumping due to the external laser. We select the value of $\eta_0$ such that the ratio of the maxima of photocurrents to the maxima of currents emerging from the standard sequential tunneling obtained in the map corresponds to the experiment. We note that the absorption of the laser photons is likely mediated by a vibronic transition. The direct excitation of the Franck-Condon $D_0^- \to D_1^-(v_i)$ transition is not considered ($\kappa_{D_1^- D_0^-} = 0$ eV) as it would result in the same physical behaviour that can be achieved by considering an internal conversion between $D_1^-$ and $D_2^-$. We consider an internal conversion $D_2^- \to D_1^-$ characterised by the rate $\hbar\kappa_{D_1^- D_2^-} = \hbar\gamma_{IC}$ = 2 meV that leads to the population of $D_2^-$ as observed in the experiment.

*The threshold function* $\Theta(E_i^{S(T)} - E_j^{S(T)})$ is obtained from a model accounting for the effect of substrate atom reorganization upon molecular charging. In essence, the substrate ions respond to the change of the perceived electric field generated by the molecule which leads to Franck-Condon effects in charge transfer from and into the molecule. In our earlier work [3] we have implemented a mathematical model for the line-shape $S(v)$ of peaks in tunneling $dI/dV$ spectra. Here we consider $S(v)$ normalized such that $\int_{-\infty}^{\infty} S(v)dv = 1$. In particular,

$$S(v) \propto Re\left\{\int_0^\infty e^{\int_0^\infty J(\Omega)e^{i\Omega t}d\Omega} e^{-ivt/\hbar} dt\right\}, \qquad (12)$$

with

$$J(\Omega) = \frac{E_R}{\left(\Omega_{max}^2 - \Omega_{min}^2\right)} rect\left[\frac{\Omega - \Omega_{min} - 0.5(\Omega_{max} - \Omega_{min})}{\Omega_{max} - \Omega_{min}}\right], \qquad (13)$$

where $\Omega_{min}$ = 18 meV and $\Omega_{max}$ = 31 meV are the lower and upper bounds of the optical-phonon frequencies in NaCl and $E_R$ = 850 meV is the reorganization energy considered in the present model, consistently with the previous work.[4] The threshold function $\Theta(E)$ is then the cumulative integral of $S(v)$

$$\Theta(E) = \int_{-\infty}^{E} S(v)dv \qquad (14)$$

Importantly, the onset of $\Theta(E)$, as compared to a regular step function, is offset due to the fact that the zero-phonon charging transition is unlikely as the reorganization energy is high. This therefore realistically accounts for the stabilization of charge states by the reorganization of the NaCl layer and is a critical component of our model as it leads to gradual variations of charge-transfer rates with varying voltage $V$. The functions $\Theta(x)$ and $S(x)$ are shown in Fig.S1.

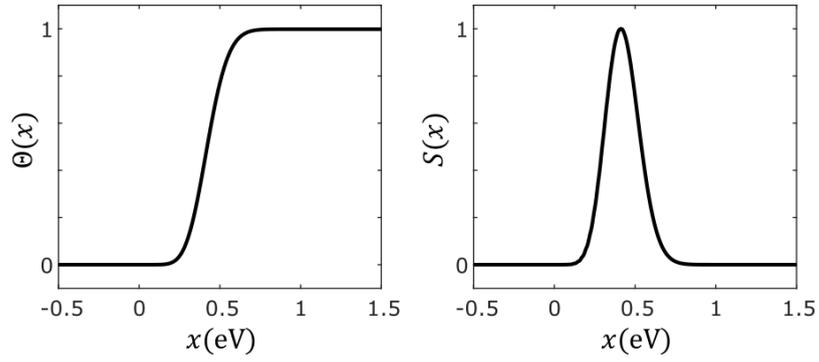

**Fig. S1: The normalized functions $\Theta(x)$ and $S(x)$ for the parameters discussed in the text.**

*Extracting many-body-state energies from the experiment.* The energies of the states considered in the model are summarized in Table 1. They have been extracted from the experimental data as follows. The vertical transition energies have been estimated from the optical emission spectra. The energies $E(S_0^{2-})$ and $E(S_0)$ have been obtained from the experimental $dI/dV$ spectra using the threshold function and assuming $\alpha = 0.8$ and $\gamma_0^T = 0.01$ μeV as the tip-sample distance is larger when the tunneling spectroscopy is performed than in the case of the photocurrent measurements.

| $E(D_0^-)$ | $E(D_1^-)$ | $E(D_2^-)$ | $E(S_0)$ | $E(S_0^{2-})$ |
|---|---|---|---|---|
| 0 | 1.330 | 1.456 | 0.200 | 0.380 |

**Table S1:** Energies of molecular states (in eV) considered in the model for $V = 0$ V. The energy of the negative ground-state (the state stabilized by the substrate at $V = 0$ V) is set to zero as a reference.

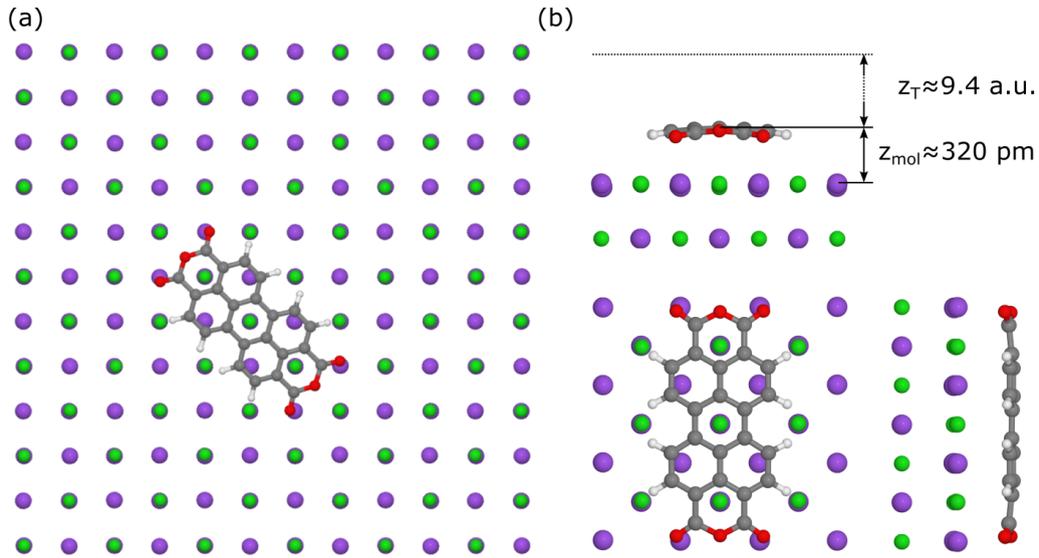

**Fig. S2: The optimized geometry of a PTCDA molecule adsorbed on a 6√2×6√2 flake of NaCl.** (a) The overview of the system and orientation of the molecule. (b) A detailed view of the adsorption geometry showing that the molecule is bent with the oxygen atoms approaching the underlying Na atoms. The distance between the highest constrained NaCl atoms and the highest point of the molecule $z_{mol}$ is shown alongside with the assumed plane where the DOS derived from the molecular orbitals is evaluated. The colors denote the atom types in the following way: Na - purple, Cl - green, O - red, C - grey and H - white.

*Ab-initio calculations.* To approximate the spatial distribution of tunneling currents and plasmon-exciton coupling we perform density-functional theory (DFT) and time-dependent DFT (TD-DFT) calculations. Since the spatial distribution of constant-height photocurrent maps can strongly depend on the details of the molecular geometry, we first optimize the molecular structure including the substrate. To that end we assume a patch of 2ML NaCl containing 6√2×6√2 unit cells as shown in Fig.S2. The patch is composed of units of 8 atoms (4 Na and 4 Cl) that form an electrostatic octupole and therefore minimize stray electric fields that could spuriously influence the properties of the molecule. Prior to relaxation we assume $d_{NaCl}$ = 2.805 Å, we then fix the positions of the atoms in the bottom layer and all the border atoms, leaving only the atoms of the top layer that have 5 nearest neighbours to relax. The molecule is placed on top of the layer as depicted in Fig.S2 consistently with experimental observations. We optimize the system geometry in the ground state assuming a singly negative doublet configuration using the hybrid B3LYP [5] functional and the 6-31G(d) double-zeta gaussian basis using the Gaussian 16 rev. C.01 software.[6] An empirical dispersive interaction (GD3BJ [7]) is considered between the atoms to account for the dispersive molecule-NaCl interaction.

After relaxing the structure, we extract the ground-state geometry of the molecule that is considerably bent compared to its flat geometry obtained in a vacuum. This geometry is used to calculate the excited states using the linear-response TDDFT implemented in Gaussian 16 and the $D_0^- \to D_1^-$ and $D_0^- \to D_1^-$ transition densities are extracted using the Multiwfn tool [8] in the form of 3D cube files that are used to calculate the spatial dependence of plasmon-exciton coupling and are shown in Fig.S3a. From the TDDFT result we also note the composition of the $D_0^- \to D_1^-$ that dominantly consists of the transitions 2 → 4 (26%) and 1→2 (74%) (see the orbital notation in Fig.2 of the main text and Fig.S3b). For definiteness, we use the information to estimate the relative weights of the two configurations contributing to the formation of $D_1^-$, although we note that a different parametrization of the model could yield similar results even if different contributions of the two configurations were considered (e.g. obtained using a different ab-initio model, perform the calculations in the excited-state molecular geometry, etc.). We note that a more precise TDDFT calculation including the substrate or relaxing the molecular geometry on the substrate in the excited-state geometry and then extracting the transition densities and other quantities is demanding and for TDDFT calculations we therefore resort to only using the ground-state relaxed geometry of the molecule.

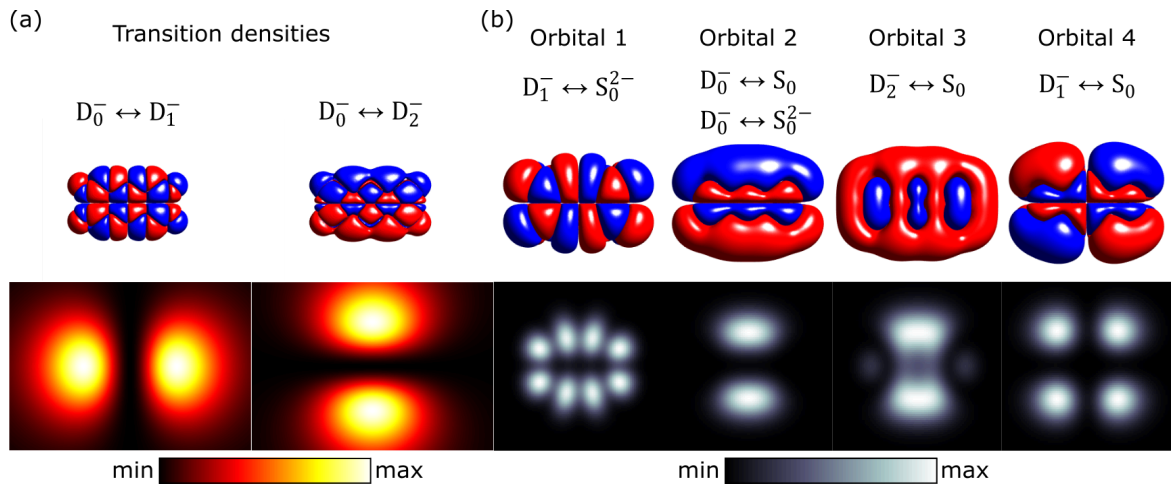

**Fig.S3: Molecular transition densities and Kohn-Sham orbitals.** (a) The $D_0^- \to D_1^-$ and $D_0^- \to D_2^-$ transition densities calculated in the ground-state geometry of the adsorbed molecule and the corresponding $\left|g_{ji}(x,y)\right|^2$ (image size 5×3.5 nm²), as described in the text. (b) The relevant Kohn-Sham molecular orbitals that we use to approximate the Dyson orbitals of the listed transitions. The DOS $\left|\psi_{ij}(x,y,z=z_T)\right|^2$ corresponding to the respective orbitals evaluated at a constant height of the tip. The spatial scale of the displayed quantities is identical and the dimensions of the DOS images are 40×40 a.u.²

Finally, to obtain the spatial dependency of the photocurrents, we approximate the Dyson orbitals of the respective transitions sketched in Fig.2. These correspond to the orbitals from which (into which) a single electron must be extracted (added) to obtain the configuration of the final state.

Since molecular orbitals constructed using gaussian basis sets do not feature a physically meaningful spatial dependence at a plane $z = z_T$ above the molecule, we resort to a different DFT software implementing the Kohn-Sham scheme on a real-space grid, Octopus v. 11.4.[9] We use the geometry obtained from Gaussian 16 and perform a ground-state DFT calculation using the Perdew-Zunger [10] parametrization of the local density approximation (LDA) correlation and the Slater density functional for the LDA exchange functional.[11,12] To achieve Kohn-Sham energies of the orbitals similar to the ionization potentials and electron affinities of the molecule, that can be estimated from Table 1 upon correction for the substrate work function ($W_{NaCl/Ag}$ ~ 3.5 eV [13]), we perform the calculation assuming a total charge of the molecule to be -0.5|$e$|. We use a tight box shape composed of atom-centered spheres of the radius 15 Å and grid step of 0.1 Å and export the orbitals into cube files. We evaluate the orbitals at a plane $z = z_T \approx 9.4$ a.u. $\approx 5$ Å above the highest point of the molecule and use these slices to define the 2D density of states (DOS) $|\psi_{ij}(x, y, z = z_T)|^2$ belonging to the respective transitions. The respective transitions and corresponding orbitals are schematically shown in Fig.2 and summarized in Fig.S3b.

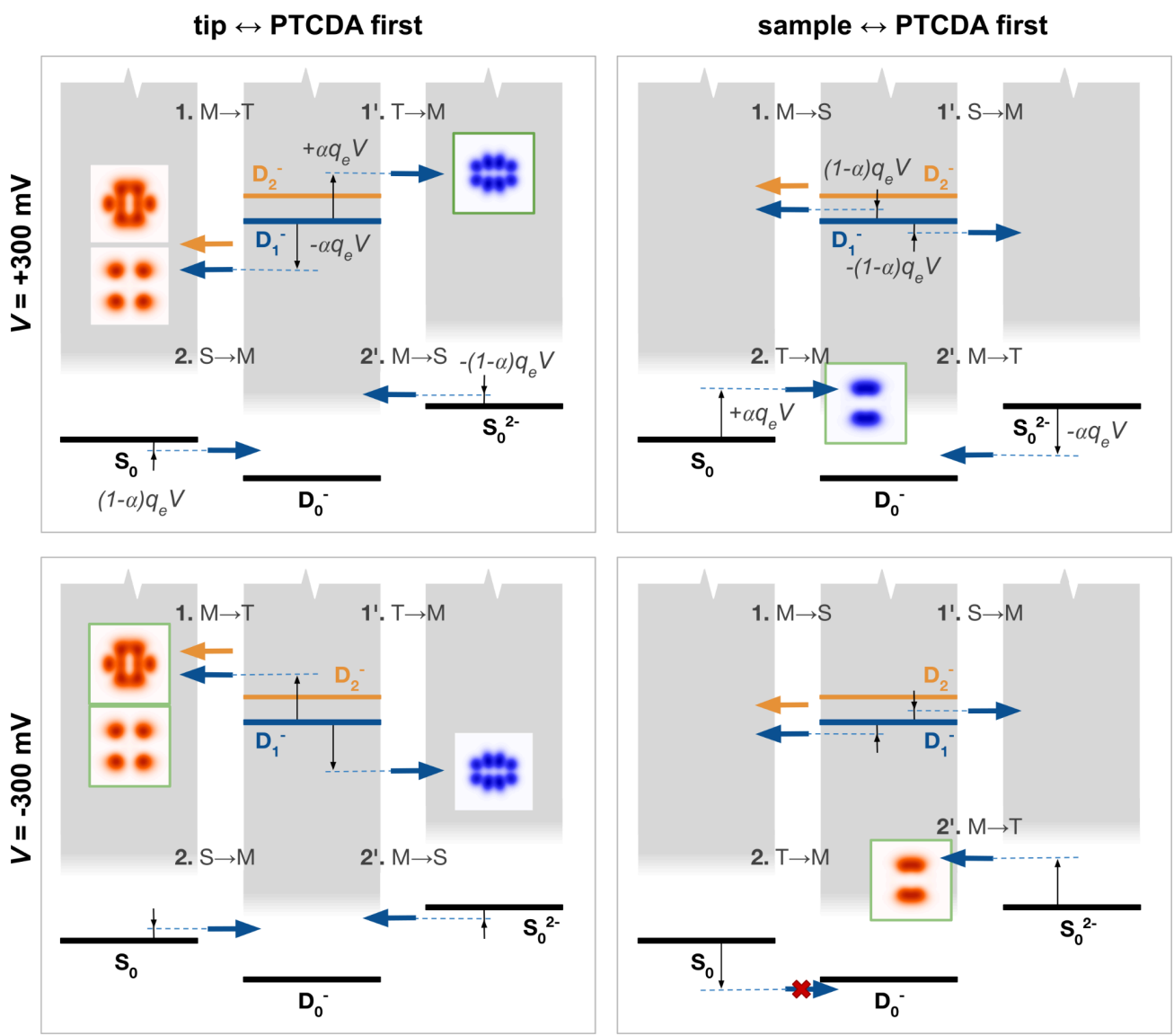

**Fig.S4: Energy schemes of the decay paths in the model with bias.** The schemes show the relative energies of the involved transitions depending on the starting transition (denoted as 1, 1', and 2, 2') and bias polarity (+300/-300 mV), together with the respective spatial modulation of the current (orange- and blue-color scale insets). The arrows indicate the energy levels of each transition at particular bias relative to the electrode involved in the transition. The gray-shaded areas represent the calculated threshold function and the center of the threshold. The order in the sequence and direction of the electron transfer is assigned to each transition; T stands for the tip, M for the molecule and S for the substrate. The Dyson-like orbitals in the insets show the tunneling probability and direction (orange and blue color scale - negative and positive net current, respectively). The orbitals which dominate the simulated photocurrent map for each of the four situations are denoted with green lining.

# Comparison of the measurements with and without irradiation to the theoretical simulations

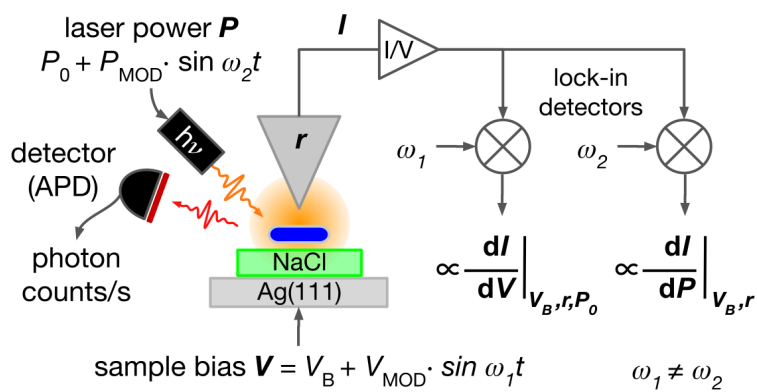

**Fig.S5: Scheme of the function of the simultaneous current, TEPL, d*I*/d*V* and d*I*/d*P* measurement.** The setup is based on standard current (*I*) and differential conductance (d*I*/d*V*) lock-in measurement with a bias modulated at a frequency below the cutoff of the high-gain *I*/*V* converter ($\omega_2$). The tip-sample cavity with the molecule is irradiated with a laser modulated at another frequency ($\omega_2$) with amplitude $P_{MOD}$ around a constant intensity level ($P_0$). The photon-induced response (photocurrent) of I is measured with a second lock-in amplifier. The two frequencies are not multiple of another. The TEPL signal is collected simultaneously with an avalanche photodiode detector. All signals are measured as a function of the relative lateral tip-sample position above the molecule.

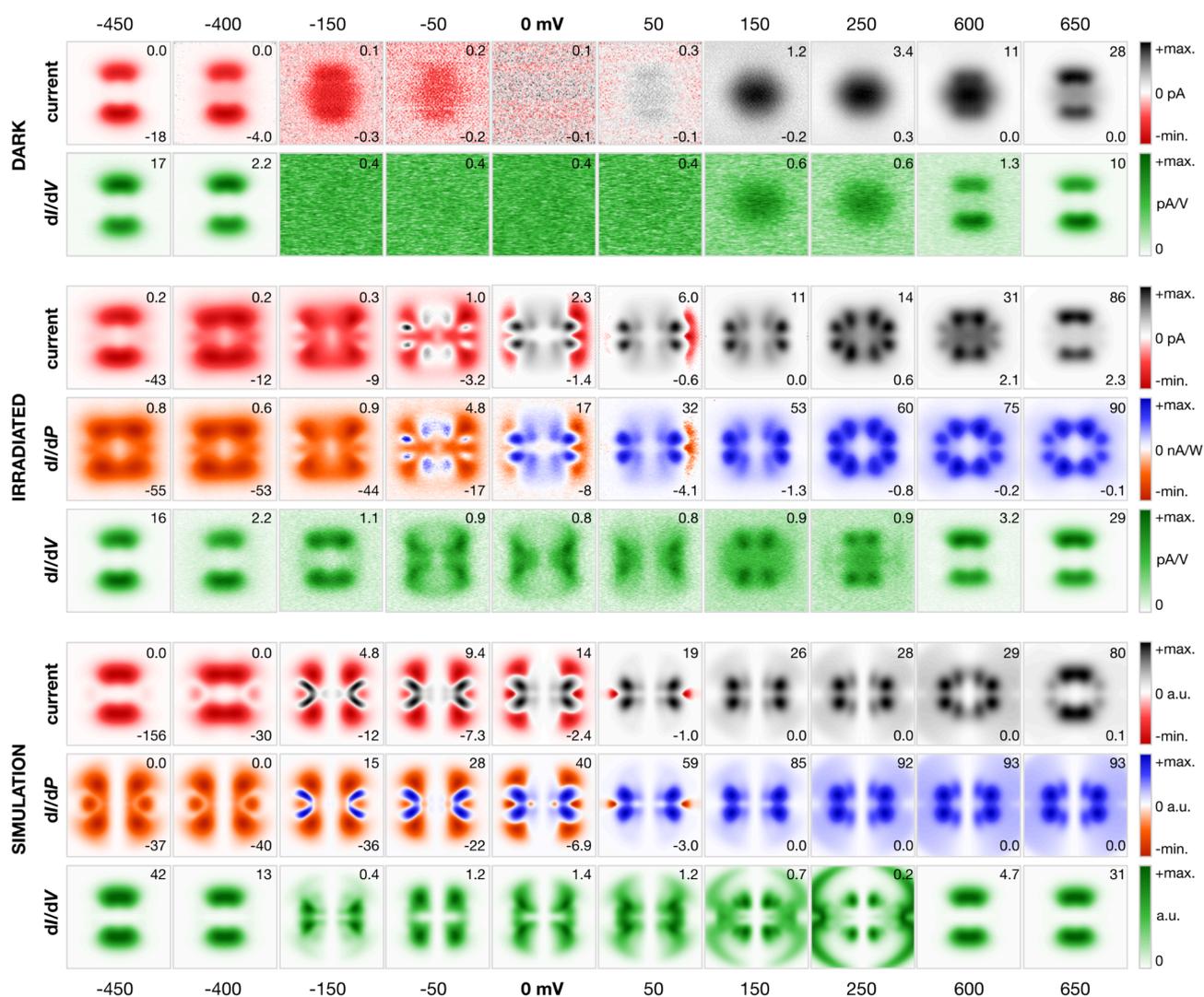

**Fig.S6: Comparison of bias- and irradiation-dependent maps of current, photocurrent, d*I*/d*V* and simulations.** The current, the photon-induced component of the current proportional to photocurrent (d*I*/d*P*) and conductance (d*I*/d*V*) maps are taken simultaneously over a PTCDA molecule in a constant-height mode as a function of the applied bias voltage in the range from -450 to 650 mV with and without the laser irradiation. The data without irradiation (dark) have been taken with a different tip, with the tip-sample distance comparable to the data taken with irradiation. Theoretical simulations are provided for comparison for all channels for the scenario with illumination. The area of the images is 2 x 2 nm$^2$. The numbers at the top and bottom right corner of each image represent the minimum and maximum data value, respectively. The "+max" and "-min" in the color scale bar refer to the maximum positive and minimum negative intensity of the signal.

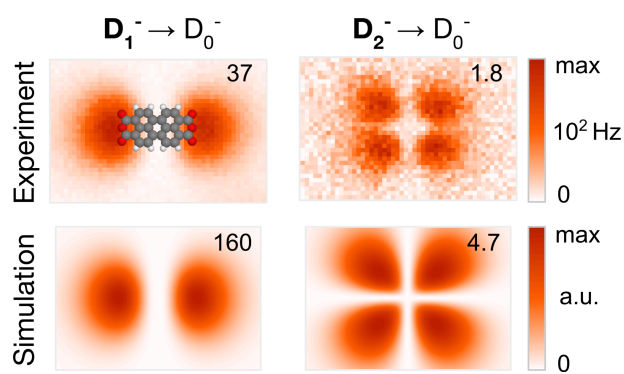

**Fig.S7: Comparison of experimental and simulated TEPL maps of the $D_1^-$ and $D_2^-$ emission at zero bias.** The experimental maps were measured by taking a TEPL spectrum in each point and plotting the intensity at the energies of the $D_1^-$ and $D_2^-$ emission peaks, integrated with a bandwidth of ~1 nm. The theoretical maps represent the rates of the $D_1^-$ and $D_2^-$ transition to the $D_0$ which are the solution of the rate equations. Dimensions of all maps are 5 x 3.5 nm$^2$.

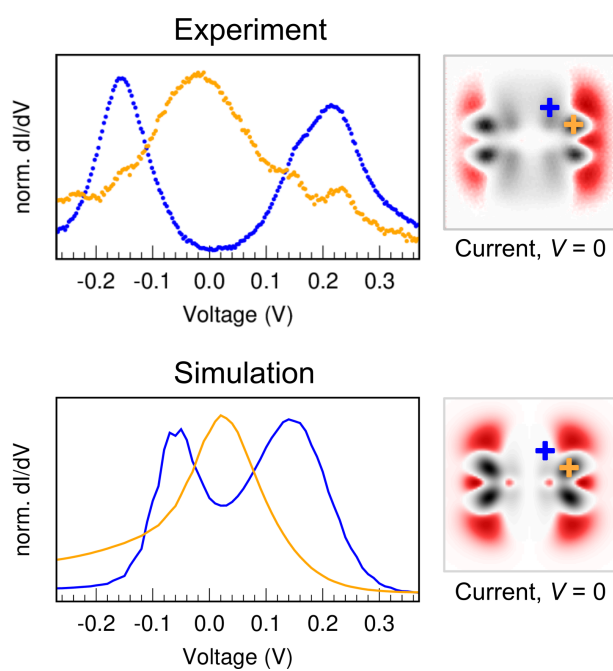

**Fig.S8: Measurement and simulation of the d$I$/d$V$ in the low-bias range.** The d$I$/d$V$ curves were measured at the two different locations above the molecule denoted by the crosses with corresponding colors. The height of the tip above the molecule corresponds to the one at which the mapping presented in Fig.S5 was performed. The simulated curves were taken in analogous locations. The bias range was set to the region between the onsets.

# References

[1] K. Kaiser, S. Jiang, M. Romeo, F. Scheurer, G. Schull, and A. Rosławska, Gating Single-Molecule Fluorescence with Electrons, Physical Review Letters 133, (2024).

[2] S. Jiang, T. Neuman, R. Bretel, A. Boeglin, F. Scheurer, E. Le Moal, and G. Schull, Many-Body Description of STM-Induced Fluorescence of Charged Molecules, Phys. Rev. Lett. 130, (2023).

[3] K. Vasilev, S. Canola, F. Scheurer, A. Boeglin, F. Lotthammer, F. Chérioux, T. Neuman, and G. Schull, Exploring the Role of Excited States' Degeneracy on Vibronic Coupling with Atomic-Scale Optics, ACS Nano 18, 28052 (2024).

[4] L. Sellies, J. Eckrich, L. Gross, A. Donarini, and J. Repp, Controlled Single-Electron Transfer Enables Time-Resolved Excited-State Spectroscopy of Individual Molecules, Nature Nanotechnology (2024).

[5] A. D. Becke, Density-functional thermochemistry. III. The role of exact exchange, The Journal of Chemical Physics 98, 5648 (1993).

[6] M. J. Frisch, G. W. Trucks, H. B. Schlegel, G. E. Scuseria, M. A. Robb, J. R. Cheeseman, G. Scalmani, V. Barone, G. A. Petersson, H. Nakatsuji, X. Li, M. Caricato, A. V. Marenich, J. Bloino, B. G. Janesko, R. Gomperts, B. Mennucci, H. P. Hratchian, J. V. Ortiz, A. F. Izmaylov, J. L. Sonnenberg, D. Williams-Young, F. Ding, F. Lipparini, F. Egidi, J. Goings, B. Peng, A. Petrone, T. Henderson, D. Ranasinghe, V. G. Zakrzewski, J. Gao, N. Rega, G. Zheng, W. Liang, M. Hada, M. Ehara, K. Toyota, R. Fukuda, J. Hasegawa, M. Ishida, T. Nakajima, Y. Honda, O. Kitao, H. Nakai, T. Vreven, K. Throssell, J. A. Montgomery, Jr., J. E. Peralta, F. Ogliaro, M. J. Bearpark, J. J. Heyd, E. N. Brothers, K. N. Kudin, V. N. Staroverov, T. A. Keith, R. Kobayashi, J. Normand, K. Raghavachari, A. P. Rendell, J. C. Burant, S. S. Iyengar, J. Tomasi, M. Cossi, J. M. Millam, M. Klene, C. Adamo, R. Cammi, J. W. Ochterski, R. L. Martin, K. Morokuma, O. Farkas, J. B. Foresman, and D. J. Fox, Gaussian 16 revision C.01 (2016).

[7] S. Grimme, S. Ehrlich, and L. Goerigk, Effect of the damping function in dispersion corrected density functional theory, J Comput Chem 32, 1456 (2011).

[8] T. Lu, A comprehensive electron wavefunction analysis toolbox for chemists, Multiwfn, The Journal of Chemical Physics 161, (2024).

[9] N. Tancogne-Dejean et al., Octopus, a computational framework for exploring light-driven phenomena and quantum dynamics in extended and finite systems, The Journal of Chemical Physics 152, (2020).

[10] J. P. Perdew and A. Zunger, Self-interaction correction to density-functional approximations for many-electron systems, Phys. Rev. B 23, 5048 (1981).

[11] P. A. M. Dirac, Note on Exchange Phenomena in the Thomas Atom, Math. Proc. Camb. Phil. Soc. 26, 376 (1930).

[12] C. Slater, A Simplification of the Hartree-Fock Method, Phys. Rev. 81, 385 (1951).

[13] M. Imai-Imada, H. Imada, K. Miwa, J. Jung, T. K. Shimizu, M. Kawai, and Y. Kim, Energy-level alignment of a single molecule on ultrathin insulating film, Phys. Rev. B 98, (2018).